\def\be{\begin{equation}}
\def\ee{\end{equation}}
\def\bea{\begin{eqnarray}}
\def\eea{\end{eqnarray}}

\def\g{\gamma}

\def\l{\lambda}

\def\n{\nu}
\def\r{\rho}

\def\p{\partial}

\def\bt{\bibitem}
\def\ct{\cite}
\documentclass[aps,pra]{revtex4}
\usepackage{graphicx}
\usepackage{shortvrb,psfrag}
\begin{document}
\title{Information Entropy and Correlation of the Hooke's Atom}
\author{Rajneesh Atre$^{1,2}$}
\email{atre@prl.ernet.in}
\author{Chandra Shekhar Mohapatra$^{3}$}
\email{chandra@phy.iitkgp.ernet.in}
\author{Prasanta K. Panigrahi$^{1}$}
\email{prasanta@prl.ernet.in} \affiliation{$^1$ Physical Research
Laboratory, Navrangpura, Ahmedabad, 380 009, India\\ $^2$ School
of Physics, University of Hyderabad, Hyderabad, 500 046, India \\
$^3$ Department of Physics, Indian Institute of Technology,
Kharagpur, 721 302, India}

\begin{abstract}
We provide an algebraic procedure to find the eigenstates of
two-charged particles in an oscillator potential, known as
{\it{Hooke's}} atom. For the planar Hooke's atom, the exact
eigenstates and single particle densities for arbitrary azimuthal
quantum number, are obtained analytically. Information entropies
associated with the wave functions for the relative motion are
then studied systematically, since the same incorporates the
effect of the Coulomb interaction. The {\it{quantum pottery}} of
the information entropy density reveals a number of intricate
structures, which differ significantly for the attractive and
repulsive cases. We indicate the procedure to obtain the
approximate eigen states. Making use of the relationship of this
dynamical system with the quasi-exactly solvable systems, one can
also develop a suitable perturbation theory, involving the Coulomb
coupling $Z$, for the approximate wave functions.
\end{abstract}
\maketitle

\section{Introduction}
Interacting charged particles in the presence of a harmonic
confinement appears naturally in a number of physical
circumstances. For example, two planar charged particles in the
presence of a constant magnetic field, interacting ions in a trap
are governed by the combined effect of Coulomb and harmonic
potentials. Unlike the Coulomb and harmonic cases, the above
artificial atom, known as {\it{Hooke's}} atom is not exactly
solvable \ct{White,Kestner,Laufer,Wagner}. The fact that, a
countable infinity of eigenstates can be identified exactly for
this case, albeit with differing oscillator frequencies, has made
Hooke's atom a very interesting test ground for checking the
effect of correlation arising due to the interaction. The model
has been used for explicating the working of density functional
theory \ct{Sahani}. The recent interest in entangled systems,
involving continuous variables, makes this model an ideal
candidate for a careful study. It should be mentioned that, the
comparably simpler system of two-interacting oscillators
\ct{Moshi} have been recently reanalyzed to see the origin of the
correlation effects, since in this case both the exact and
Hartree-Fock solutions are known \ct{March}.

After perturbative and numerical investigations of the Hookean
atom \ct{White,Kestner,Laufer,Wagner}, Kais {\em{et al.,}} were
able to identify one state analytically for a specific value of
the spring constant \ct{Kais}. The same system, with an additional
linear term in the inter-particle potential, was also analyzed by
Ghosh {\em{et al.,}} to study the single particle density
\ct{Ghosh}. Later on, Taut studied this system carefully and
through the analysis of the three term recurrence relation showed
that, the system possesses analytical solutions for a particular,
infinitely denumerable set of oscillator frequencies
\ct{Taut,Taut1,Turb}. He further showed that, two conditions need
to be satisfied for obtaining the above solutions. The first one
relates the energy with the oscillator frequency and other quantum
numbers of the system. The general expression for the second
condition, which connects oscillator frequency with the Coulomb
coupling and other quantum numbers have not been obtained exactly;
the solutions for the first few values have been worked out.
Recently Holas {\em{et al.,}} have studied the general structure
of the density matrix for this system, concentrating on a single
value of the spring constant earlier considered by Kais {\em{et
al}} \ct{March1}.

In this paper, we provide a general algebraic procedure to find
the eigenstates of planar Hookean atom, which can be extended to
higher dimensions. Our procedure allows one to compute the series
expression of the wave function explicitly in a more economic
manner as compared to the routinely used series solution approach.
It is based on a novel method for obtaining the solutions of
linear differential equations \ct{pkp,charan}. Apart from
obtaining the analytical expressions, in the present approach one
can also carry out a perturbative expansion for the wave function,
when closed form solutions are not available. The analytical
expressions are then used to compute the exact single particle
densities for arbitrary values of the quantum numbers. The
densities and the information entropy, associated with the pair
correlation function are systematically studied to analyze the
effect of the interaction on correlation in this artificial atom.

In the following section, we outline the above mentioned method
for solving linear differential equations and employ the same to
systematically study the eigenstates of the Hooke's atom. The
single particle densities for the planar Hooke's atom  for
arbitrary values of azimuthal quantum number are obtained
analytically. The single particle densities and information
entropy, associated with the relative motion, are then analyzed.
The effect of interaction on correlation is shown to be revealed
by information entropy. We then point out, how the connection of
Hooke's atom with the well-studied quasi-exactly solvable (QES)
system can be used to develop perturbative expansion involving
Coulomb coupling parameter $Z$. We conclude in section III after
pointing out directions for future studies.

\section{Hooke's Atom} The Hooke's atom, dealing with two
interacting charged particles in harmonic potential, is governed
by the Hamiltonian (in units, $\hbar=m=e=1$):

\be \label{HK2D} H = \sum_{i=1}^{2} \left \{ \frac{1}{2}
\left({\bf{p}}_{i}+\frac{1}{c} {\bf{A(r_i)}}
\right)^{2}+\frac{1}{2}\omega_{0}^{2}r_{i}^{2}
\right\}+\frac{Z}{|{\bf{r}}_2 -{\bf{r}_1|}}. \ee

This Hamiltonian decouples in the center of mass
${\bf{R}}=\frac{1}{2}({\bf{r}}_1+{\bf{r}}_2)$ and the relative
coordinate, ${\bf{r}}={\bf{r}}_2-{\bf{r}}_1$, which give rise to
the new momentum operators,
${{\bf{p}}=-i\bf{\nabla}}_{r}=\frac{1}{2}\left({\bf{p}}_{2}-{\bf{p}}_{1}\right)$
and ${\bf{P}}=-i{\bf{\nabla}}_{R}={\bf{p}}_{2}+{\bf{p}}_{1}$.

Eq.(\ref{HK2D}) in these coordinates reads

\begin{widetext}
\bea \label{HK2D_rel} H&=&2\left\{\frac{1}{2}\left[
{\bf{p}}+\frac{1}{c}{\bf{A_{r}}}\right]^{2}+\frac{1}{2}\omega_{r}r^{2}+\frac{Z}{2r}
\right\}+\frac{1}{2}\left\{\frac{1}{2}\left[
{\bf{P}}+\frac{1}{c}{\bf{A_{R}}}\right]^{2}+\frac{1}{2}\omega_{R}R^{2}
\right\}~, \\\label{H2KSpl} &\equiv& 2H_r+\frac{1}{2}H_R \eea
\end{widetext}
where
${\bf{A_R}}=2{\bf{A(R)}}={\bf{A}(\bf{r}_1)}+{\bf{A}(\bf{r}_2)}$,
${\bf{A_r}}=\frac{1}{2}{\bf{A(r)}}=\frac{1}{2}({\bf{A}(\bf{r}_2)}-{\bf{A}(\bf{r}_1)})$
and $\omega_{R}=2\omega_{0}$, $\omega_{r}=\frac{1}{2}\omega_{0}$.
The total wave function factorizes: \be
\Psi(1,2)=\xi({\bf{R}})\varphi({\bf{r}})~.\ee

It is clear from Eq.(\ref{H2KSpl}) that, the eigenstates of $H_R$
are identical with the two dimensional oscillator states, where
${\bf{A_R}}$ is a linear function of $R$, for a constant magnetic
field $B$.

The Schr\"{o}dinger equation $H_{r}\varphi({\bf{r}})=\varepsilon
\varphi({\bf{r}})$ for the internal motion, can be cast in the
form,
\be \left\{-\frac{1}{2}\left[r^{-1/2}\frac{\p^{2}}{\p
r^2}r^{1/2}+\frac{1}{r^2}\left(\frac{\p^{2}}{\p\phi^2}+\frac{1}{4}
\right)\right]-\frac{i
\omega_{L}}{2}\frac{\p}{\p\phi}+\frac{1}{2}\left[\omega_{r}^{2}+\frac{1}{4}\omega_{L}^{2}\right]r^{2}+
\frac{Z}{2r}\right\}\varphi(\bf{r})=\varepsilon \varphi(\bf{r}).
\ee Here $\omega_L$ is the Larmor frequency. The angular and
radial part of the wave function $\varphi(\bf{r})$, are decoupled
through the ansatz,
\be\varphi({\bf{r}})=\frac{e^{im\phi}}{\sqrt{2\pi}}\frac{u(r)}{r^{1/2}}~~~~~
m=0,\pm1,\pm2,\cdots, \ee where $u(r)$ satisfies the radial
equation,
\begin{widetext} \be
\left\{-\frac{1}{2}\frac{d^2}{dr^2}+\frac{1}{2}\left(m^2
-\frac{1}{4}\right)\frac{1}{r^2}+\frac{1}{2}\tilde{\omega}_{r}^{2}r^2+\frac{Z}{2r}
\right\}u(r)=\left[\varepsilon-\frac{1}{2}m\omega_{L}\right]u(r).
\ee Here
$\tilde{\omega}_r=\frac{1}{2}\sqrt{\omega_{L}^{2}+\omega_{0}^{2}}$.
\end{widetext}
To solve the above equation, one introduces the dimensionless
variable $\rho =\sqrt{\tilde{\omega}_{r}}~r$ and
\be\varepsilon'=\frac{2}{\tilde{\omega}_r}\left(\varepsilon-\frac{1}{2}m\omega_L\right)~,\ee
and substitute \be
u=e^{-\frac{1}{2}\rho^2}\r^{|m|+1/2}t(\rho)~.\ee This yields the
following equation:

\be \label{RadReq} \left\{
\frac{d^2}{d\r^2}-2\r\frac{d}{d\r}+2\left(|m|+\frac{1}{2}\right)\frac{1}{\r}\frac{d}{d\r}
-\frac{Z}{\sqrt{\tilde{\omega}_{r}}\r}
+\left(\varepsilon'-2|m|-2\right)\right\}t(\r)=0 ~.\ee

The above radial equation can be solved for a series solution,
following a recently developed method for solving linear
differential equations \ct{pkp,charan}. A single variable
differential equation, after suitable manipulations, can be cast
in the form,

\be \left[F(D)+P(x,d/dx)\right]y(x)=0~~; \ee here $F(D)$ is a
function of the Euler operator $D\equiv x\frac{d}{dx}$, possibly
containing a constant. $P(x,d/dx)$ contains other operators of the
differential equation concerned. The series solution for $y(x)$
can be written in the form:

\be\label{formal} y(x)=
C_{\l}\left\{\sum_{\nu=0}^{\infty}(-1)^\n\left[\frac{1}{F(D)}P(x,d/dx)\right
]^{\n}\right\}x^\l,\ee provided $F(D)x^{\l}=0$.

The proof of this can be be checked by direct substitution. For
the Hooke's atom case, we have $F(D) \equiv D(D+2m)$, $\l=0, -2m$
and $P \equiv \left(\varepsilon'-2|m|-2\right)\r^2-2\r^3
\frac{d}{d\r}-\frac{Z\r} {\sqrt{\tilde{\omega}_{r}}}$, after
multiplying $\r^2$ to the Eq.(\ref{RadReq}). Eq.(\ref{formal})
yields, \be t(\r)=
C_{0}\left\{\sum_{\nu=0}^{\infty}(-1)^\n\left[\frac{1}{D(D+2m)}\left\{\left(\varepsilon'-2|m|-2\right)\r^2-2\r^3
\frac{d}{d\r}-\frac{Z\r} {\sqrt{\tilde{\omega}_{r}}}\right\}\right
]^{\n}\right\}\r^0~, \ee which can be expanded in the form, \bea
\label{sol}
t(\r)&=&1+\frac{Z}{{\sqrt{\tilde{\omega}_{r}}}(2m+1)}\r+
\left(\frac{Z^2}{\tilde{\omega}_{r}(2m+1)2(2m+2)}-\frac{\tilde{E}}{2(2m+2)}\right)\r^2
\nonumber
\\
&+&
\frac{1}{\sqrt{\tilde{\omega}}_{r}}\left\{\frac{Z^3}{\tilde{\omega}_{r}(2m+1)2(2m+2)3(2m+3)}-
\frac{Z\tilde{E}}{(2m+1)3(2m+3)} \right.\nonumber
\\&+& \left. \frac{Z\tilde{E}}{2(2m+2)3(2m+3)}- \frac{2Z}{(2m+1)3(2m+3)}
\right\}\r^3+\cdots ~~,
~~{\mathrm{here}}~~\tilde{E}=\varepsilon'-2(|m|+1)~. \eea In order
that, the above series terminates at $\r^{n-1}$, the coefficients
of $\r^n$ and $\r^{n+1}$ terms must be zero. There are two
operators $\left\{\left(\varepsilon'-2|m|-2\right)\r^2-2\r^3
\frac{d}{d\r}\right\}$ and
$\frac{Z\r}{\sqrt{\tilde{\omega}_{r}}}$, which increase the degree
of the polynomial solution. Once $a_{n}$, the coefficient of
$\r^n$ is zero, $\r^{n+1}$ term will not get any contribution from
$\frac{Z\r}{\sqrt{\tilde{\omega}_{r}}}$ and the only contribution
to $a_{n+1}$ will arise from the action of
$\left\{\left(\varepsilon'-2|m|-2\right)\r^2-2\r^3
\frac{d}{d\r}\right\}$ on $a_{n-1}\r^{n-1}$. Hence, $a_{n+1}=0$
yields, $\varepsilon'= 2(n+|m|)$. The coefficients $a_{n}$ can be
straightforwardly determined in our approach; although it becomes
tedious for higher values of $n$. Once the above condition is
implemented, one needs to check if $a_{n}=0$ yields exact
solutions.

 For $n=2$ and arbitrary $m$ and $Z$, we have the frequencies
$$ \tilde{\omega}_r=\frac{Z^2}{2(2|m|+1)}~~
{\mathrm{and~~energy}}~~ \varepsilon=\frac{Z^2
(|m|+2)}{(2|m|+1)}+\frac{1}{2}m\omega_L$$ and
\be
u(r)=1+\frac{Zr}{(2|m|+1)}. \ee

For $n=3$ and arbitrary $m$ and $Z$, we have the frequencies
$$ \tilde{\omega}_r=\frac{Z^2}{4(4|m|+3)}~~
{\mathrm{and~~energy}}~~ \varepsilon=\frac{Z^2
(|m|+2)}{2(4|m|+3)}+\frac{1}{2}m\omega_L$$ and
\be
u(r)=1+\frac{Zr}{(2|m|+1)}+\frac{Z^2r^2}{2(2m+1)(4m+3)}. \ee

In case of $n=4$, we have two roots for the oscillator frequency:
$$\tilde{\omega}_r=\frac{10Z^2+10Z^2|m|\pm
Z^2{\sqrt{73+128|m|+64|m|^2}}}{18(4|m|^2+8|m|+3)}~~.
$$ For the lower frequency the polynomial part of the wave function is given by,

\bea
u(r)&=&1+\frac{Zr}{(2|m|+1)}+\frac{Z^2\left({\sqrt{73+128|m|+64|m|^2}}-4|m|-1\right)}{12(|m|+1)(4|m|^2+8|m|+3)}r^2
\nonumber
\\&+&\frac{Z^3\left(11{\sqrt{73+128|m|+64|m|^2}}+2|m|(7{\sqrt{73+128|m|+64|m|^2}}-89)-104|m|^2-83\right)}
{108(|m|+1)(4|m|^2+8|m|)^2}r^3, \eea with energy
$\varepsilon=2(n+|m|){\tilde{\omega}}_{r}$. The other eigenvalues
and eigenstates can be found analogously. We note that, for the
general case the coefficients of $\r^{n+1}$ yields a relation
between the energy, frequency and the azimuthal quantum number. It
is the second relationship {\em{i.e.,}} $a_n=0$ which brings the
relationship between $Z$, $\tilde{\omega}_r$ and $m$. It is worth
mentioning that, using the above series expansion, it is possible
to approximately determine the other analytically inaccessible
states of a given Hamiltonian. Although, we intend to deal with
this in greater detail later, below we briefly outline the
underlying procedure. To be specific, consider the repulsive $n=2$
case, for which the ground state is analytically available for a
particular frequency. To obtain approximately, the first excited
state of the same Hamiltonian, one starts with the above series
expansion, retaining sufficient number of terms to ensure that the
wave function has one real zero in the half line. In this wave
function the energy can be treated as a variational parameter, to
accurately determine the eigenvalue and eigen state, as has been
shown for the QES sextic oscillator previously \ct{Atre}. This
process will be made more transparent while demonstrating the
connection of the Hooke's atom with the above mentioned sextic
oscillator problem below.

We now proceed to compute the pair correlation function and single
particle density defined respectively as,

\be G({\bf{r}})= \left\langle \Psi{\left|\sum_{i<j}
\delta({\bf{r}}_{i}-{\bf{r}}_{j}-{\bf{r}})\right|} \Psi
\right\rangle=|\varphi({\bf{r}})|^{2}, \ee

\be n({\bf{r}})= \left\langle \Psi{\left|\sum_{i=1}^{2}
\delta({\bf{r}}-{\bf{r}}_{i})\right|} \Psi \right\rangle~. \ee

The expression for the charge density  for $n=2$, $m=0$ and $Z=1$
can be computed:\begin{widetext} \bea
n({\bf{r}})=&~& \frac{1}{125 (3+\sqrt{2\pi})} \nonumber \\
&\times&
\left[16e^{-\frac{9r^2}{20}}\pi^3\left\{e^{\frac{r^2}{20}}\left(65+4r^2\right)+{\sqrt{10\pi}\left(10
+r^2\right)I_{0}\left(\frac{r^2}{20}\right)}+{\sqrt{10\pi}}r^2I_{1}\left(\frac{r^2}{20}\right)
\right\} \right] ~~.\eea
\end{widetext}

The corresponding attractive case ($Z=-1$) is given by:

\bea n({\bf{r}})=&~& \frac{1}{125 (3-\sqrt{2\pi})} \nonumber \\
&\times&
\left[16e^{-\frac{9r^2}{20}}\pi^3\left\{e^{\frac{r^2}{20}}\left(65+4r^2\right)-{\sqrt{10\pi}\left(10
+r^2\right)I_{0}\left(\frac{r^2}{20}\right)}-{\sqrt{10\pi}}r^2I_{1}\left(\frac{r^2}{20}\right)
\right\} \right] ~~.\eea

Similarly for the repulsive case, with $m=1$, we have

\bea \nonumber n({\bf{r}})=\frac{1}{9375(42+9\sqrt{6\pi})}
\left[64e^{-\frac{3r^2}{20}}\pi^3\left\{e^{\frac{r^2}{60}}\left(13950+705r^2+4r^4\right)
\right. \right.
\\
\left. \left. +{\sqrt{30\pi}\left(1350+90r^2+r^4\right)I_{0}
\left(\frac{r^2}{60}\right)}+{\sqrt{30\pi}}r^2\left(60+r^2\right)I_{1}
\left(\frac{r^2}{60}\right)\right\} \right].\eea

For $n=3$, $Z=1$ and $m=0$,

\bea \nonumber n({\bf{r}})=\frac{1}{28125(25+8\sqrt{3\pi})}
\left[16e^{-\frac{3r^2}{40}}\pi^3\left\{e^{\frac{r^2}{120}}\left(106425+2160r^2+4r^4\right)
\right.\right. \\
\left. \left. -{\sqrt{15\pi}\left(15300
+435r^2+2r^4\right)I_{0}\left(\frac{r^2}{120}\right)}-{\sqrt{15\pi}}r^2\left(315+2r^2\right)I_{1}
\left(\frac{r^2}{120}\right) \right\} \right] ~~.\eea

We give below the three dimensional plots of information entropy
densities associated with the pair correlation function
$G({\bf{r}})=|\varphi({\bf{r}})|^{2}$, as the same incorporates
the interaction, defined as
$S_{G}({\bf{r}})=-G({\bf{r}})\ln[G({\bf{r}})]$, for various values
of $n$ and Coulomb couplings. The effect of interaction on the
correlation is clearly seen in these plots. In Fig.1 we have
plotted $S_{G}({\bf{r}})$ for $n=2$, $Z=-1$ and various values of
azimuthal quantum number $m=0,1$ and $2$. We observe a dip at the
origin; this occurs because $G({\bf{r}})>1$ at $r=0$ which makes
the quantity $S_{G}({\bf{r}})<0$. This is a special case, as for
higher $n$ values and $m=0$, we observe a peak at the origin,
since $G({\bf{r}})<1$. For higher values of $m$ correlation
decreases and wave function of the system delocalizes. Similar
behavior is seen in Fig.2. Fig. 3 explicates the effect of Coulomb
interaction on the entropy density; Fig.3 (a), (b) and (c) are
plotted for various coupling strengths keeping $n$ and $m$ fixed.
It is clear that, as the coupling strength increases, the system
becomes more localized. In Fig. 4, we have plotted the entropy
density for repulsive Hooke's atom with $n=4$. We observe, for the
$m=0$ there is a small dip at the center, which becomes wider for
higher values of $m$, which agrees with the fact that for higher
values of $m$ there is less correlation present in the system.

In Fig. 5, we have numerically calculated the position space
entropy for various values of $m$, keeping $n=3$, for attractive
and repulsive cases. It is clear from the plot that, as the
oscillator frequency increases (which corresponds to a smaller
value of $m$), entropy decreases. Also an attractive system has
more entropy as compared to its repulsive counterpart.


The Hooke's atom can be connected to the quasi-exactly solvable
models by a simple coordinate transformation. We can make use of
this connection to develop a suitable perturbation theory
involving the coupling parameter $Z$. The QES systems are
intermediate to exactly solvable and non-solvable systems. Unlike
the exactly solvable models, in such systems only a finite number
of states can be determined analytically \ct{Shifm,Ushve}. The
typical example of a QES system, which we are going to map with
Hooke's atom is a {\em{sextic oscillator}} with a centrifugal
barrier, whose Hamiltonian is given by,

\begin{equation} \label{QESH}
\hat{H}=-\frac{1}{2}\frac{d^2}{dx^2}+\frac{\alpha x^2
}{2}+\frac{\gamma x^6}{2}+\frac{m(m+1)}{2x^2}~~.
\end{equation}

We make the following similarity transformation
$\tilde{H}=\psi_{0}^{-1}(x)\hat{H}\psi_{0}(x)$, with
$\psi_{0}(x)=x^{m+1}\exp(-x^4{\sqrt{\gamma}}/4)$ and seek the
polynomial solution for the reduced Hamiltonian,

\begin{widetext}
\be \label{qes-red}
\tilde{H}=-\frac{1}{2}\frac{d^2}{dx^2}+\sqrt{\gamma}x^{3}\frac{d}{dx}+
\left(\frac{\alpha}{2}+\frac{3\sqrt\gamma}{2}+(m+1)\sqrt{\gamma}\right)x^2
- (m+1)\frac{1}{x}\frac{d}{dx}~~. \ee
\end{widetext}
The series solution of the above equation, in the present
approach, can be written for $F(D)\equiv D(D+2m+1)$:

\bea
u(x)&=&C_{0}\sum_{\n=0}^{\infty}\left\{(-1)^{\n}\left[\frac{1}{F(D)}
\left(Ex^2-Ax^4-2\sqrt{\g}x^5\frac{d}{dx}\right)\right]^{\n}\right\}x^0,
\\
u(x)&=&1-\frac{E}{2(2m+3)}x^2+\left\{\frac{A}{4(2m+5)}+\frac{E^2}{2(2m+3)4(2m+5)}
 \right\}x^4\nonumber\\&-&\left\{\frac{E^3}{2(2m+3)4(2m+5)6(2m+7)}+\frac{EA}{4(2m+5)6(2m+7)}
+\frac{EA}{2(2m+3)6(2m+7)} +\frac{4E\sqrt{\g}}{2(2m+3)6(2m+7)}
\right\}x^6+\cdots ~,\eea where,
$A=\left(\frac{\alpha}{2}+\frac{3\sqrt\gamma}{2}+(m+1)\sqrt{\gamma}\right).$

It is worth noting that, the operators in the above equation,
$\sqrt{\gamma}x^{3}\frac{d}{dx}$ and
$\left(\frac{\alpha}{2}+\frac{3\sqrt\gamma}{2}+(m+1)\sqrt{\gamma}\right)x^2$
increase the degree of the polynomial by two. In order to preserve
the degree of the polynomial in an eigenvalue equation of
$\tilde{H}$, the following condition on the parameters should be
imposed, \be \label{condi} \frac{\alpha}{2}+ \frac{3\sqrt
\gamma}{2}+(m+1)\sqrt\gamma=-n \sqrt \gamma ~~.\ee

A simple transformation of variable: $x^2=r$ in Eq.(\ref{QESH})
allows us to map this QES problem to Hooke's atom;
\begin{widetext}
\bea \left[-\frac{1}{2} \left\{
4r\frac{d^2}{dr^2}+2\frac{d}{dr}\right\}+\frac{\alpha
r}{2}+\frac{\gamma r^3}{2}+\frac{m(m+1)}{2r} \right]\psi(r)=E
\psi(r)~~, \nonumber \\ {\mathrm{or}}~~
\left[-\frac{d^2}{dr^2}-\frac{1}{2r}\frac{d}{dr}+\frac{\gamma}{4}
r^2+\frac{m(m+1)}{4r^2}-\frac{E}{2r}
\right]\psi(r)=-\frac{\alpha}{4}\psi(r)~~. \eea
\end{widetext}
With $\psi(r)=r^{-1/4}\varphi(r)$ and
$(\tilde{m}^{2}-1/4)=\frac{m(m+1)}{4}-\frac{3}{16}$ the above
equation, comes to the form of Hooke's atom,

\begin{equation}
\left[-\frac{1}{2}\frac{d^2}{dr^2}+\frac{\gamma}{8}r^{2}+\frac{1}{2}\left(\tilde{m}^{2}-\frac{1}{4}\right)
\frac{1}{r^2}-\frac{E}{4r}\right]\psi(r)=-\frac{\alpha}{8}\psi(r).
\end{equation}
It is interesting to notice that, energy eigenvalue, quadratic and
sextic couplings of the QES Hamiltonian become the Coulomb
coupling, energy and quadratic coupling of the corresponding
Hookean system respectively, but the centrifugal barrier term
remains same in both the cases. With this parameterization, it can
be easily seen that the QES condition given by Eq.(\ref{condi}) is
same as the reduced energy $\varepsilon'=2(n+|m|)$. As has been
mentioned earlier, for finding the analytically inaccessible QES
states, one can start with the above series expansion and
terminate it at a desired degree, depending on the required number
of the nodes of the concerned wave function. Treating energy as a
variational parameter, one minimizes $\langle
\Psi|(H-E)|\Psi\rangle$, to obtain an accurate expression for the
eigen state and eigen value \ct{Atre}. The perturbation in energy
in QES regime implies a perturbation in $Z$ in its Hookean
counterpart.

\begin{figure*}
\begin{tabular}{ccc}
\scalebox{0.45}{\includegraphics{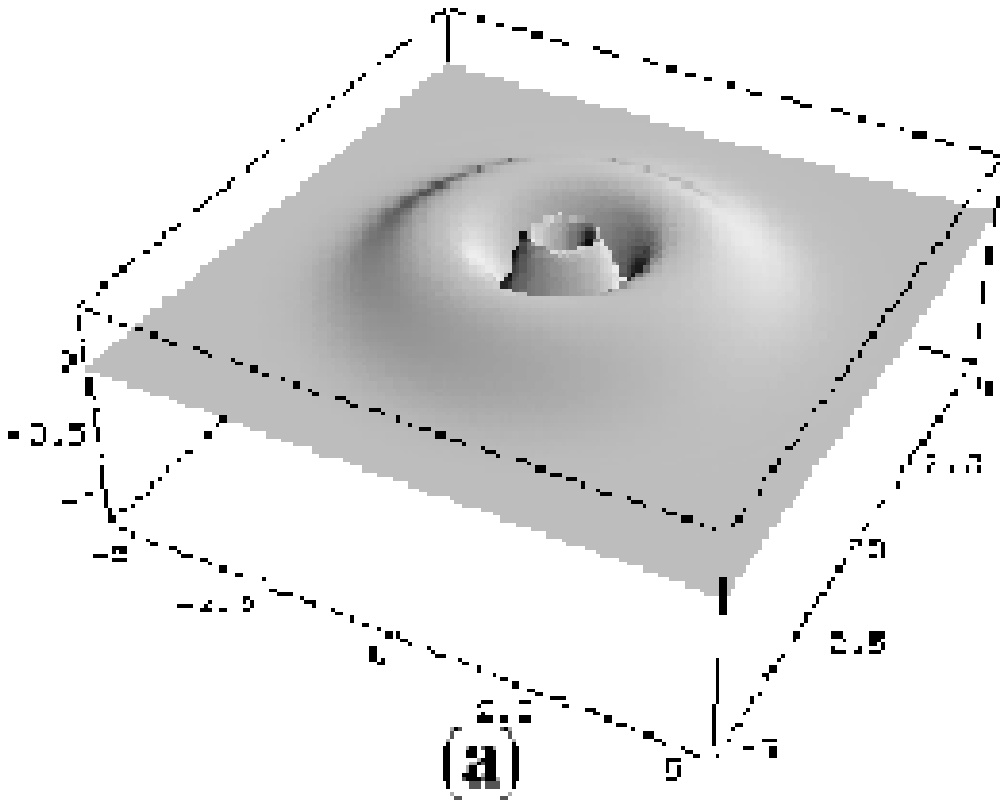}}&
\scalebox{0.45}{\includegraphics{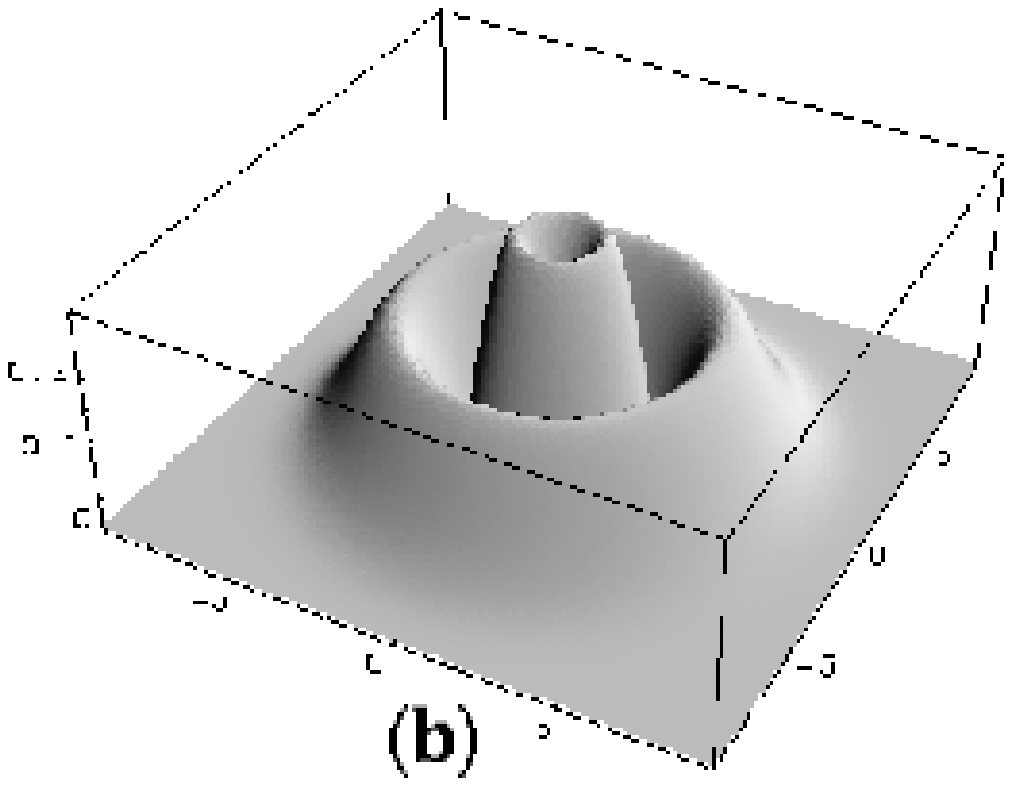}}&
\scalebox{0.45}{\includegraphics{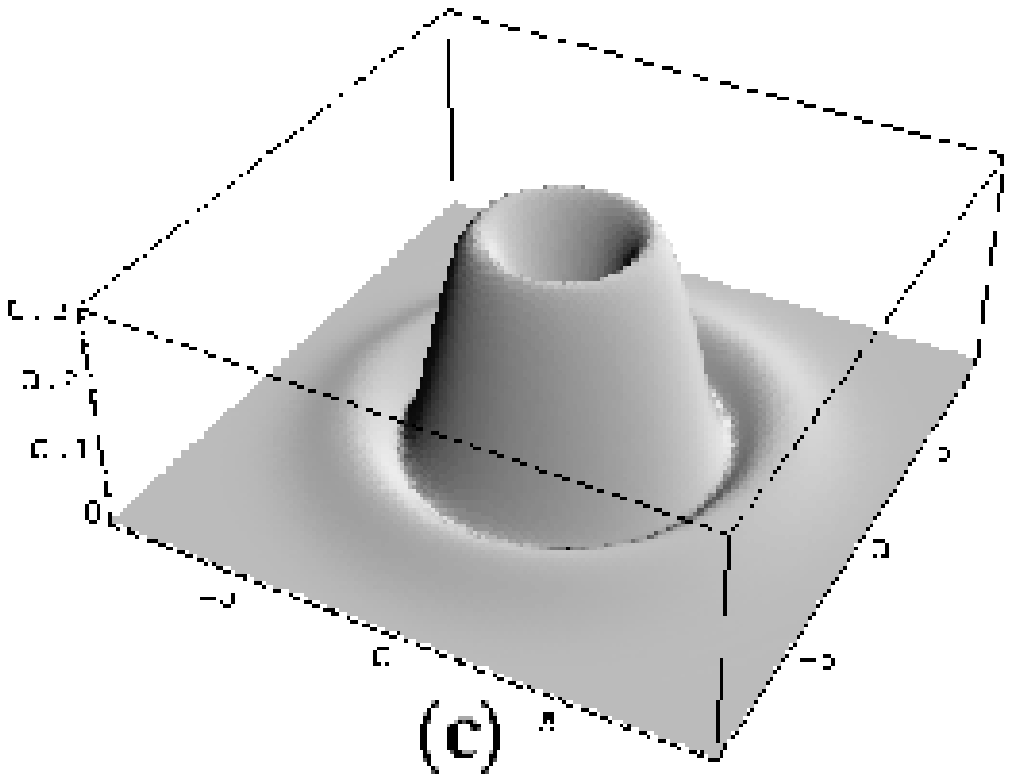}}
\end{tabular}
\caption{\label{SED}Plots of information entropy density for
attractive $Z=-1$ planar Hooke's atom with $n=2$ and  (a) $m=0$,
(b) $m=1$ and (c) $m=2$.}
\end{figure*}

\begin{figure*}
\begin{tabular}{ccc}
\scalebox{0.45}{\includegraphics{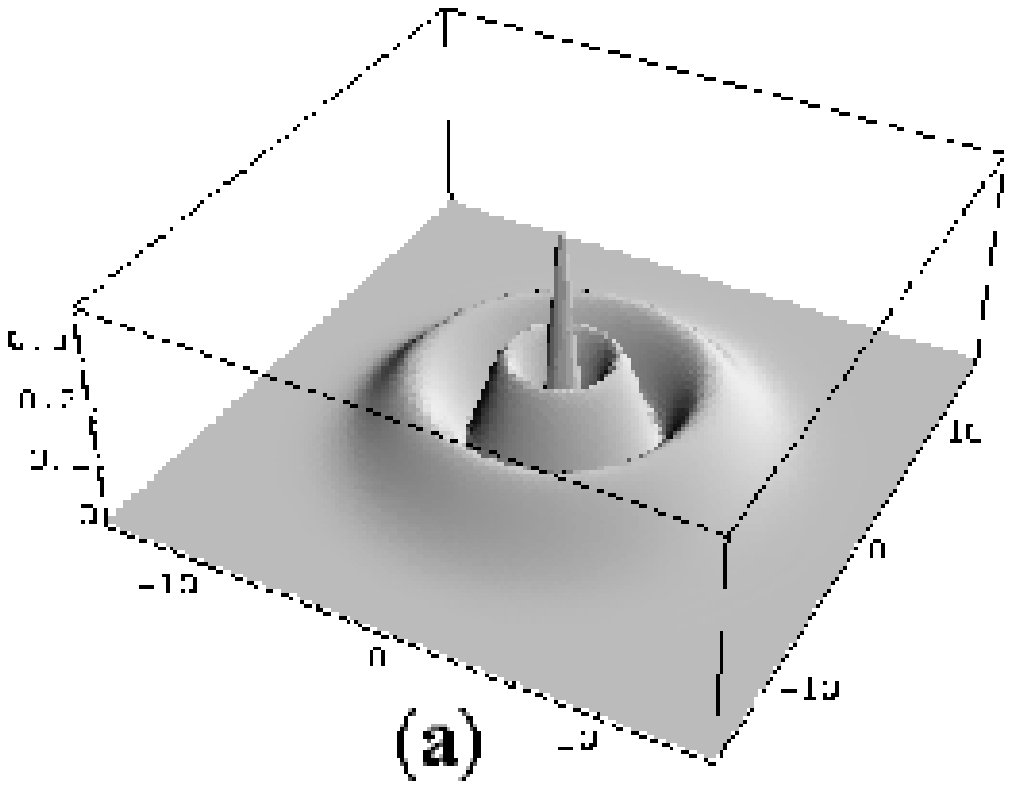}}&
\scalebox{0.45}{\includegraphics{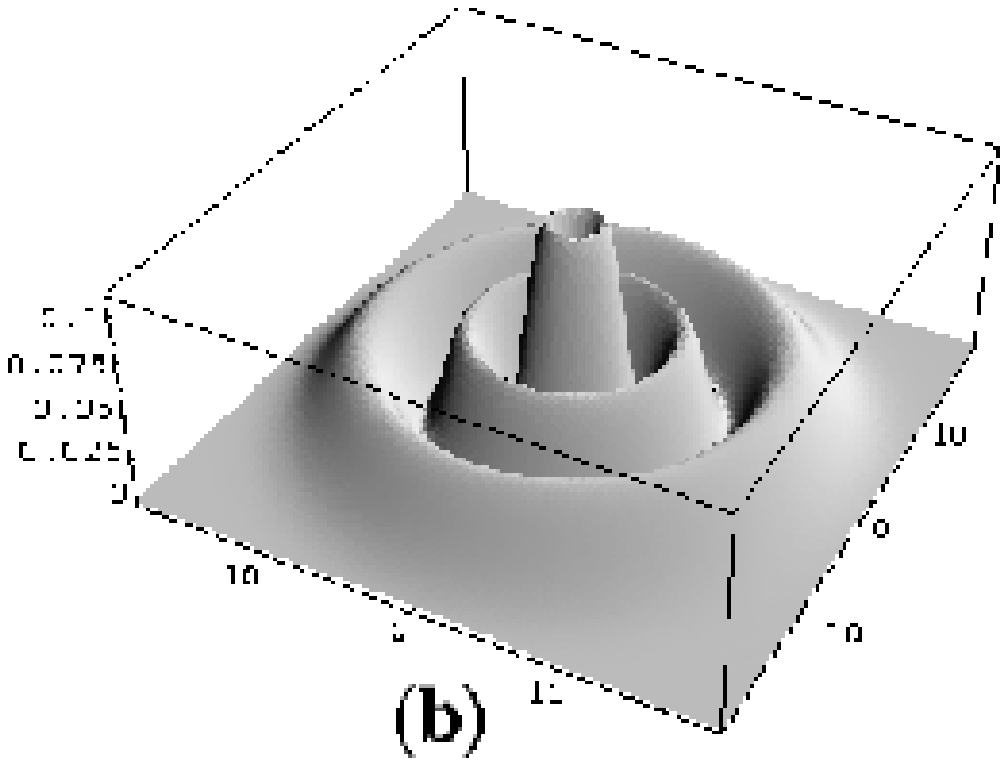}}&
\scalebox{0.45}{\includegraphics{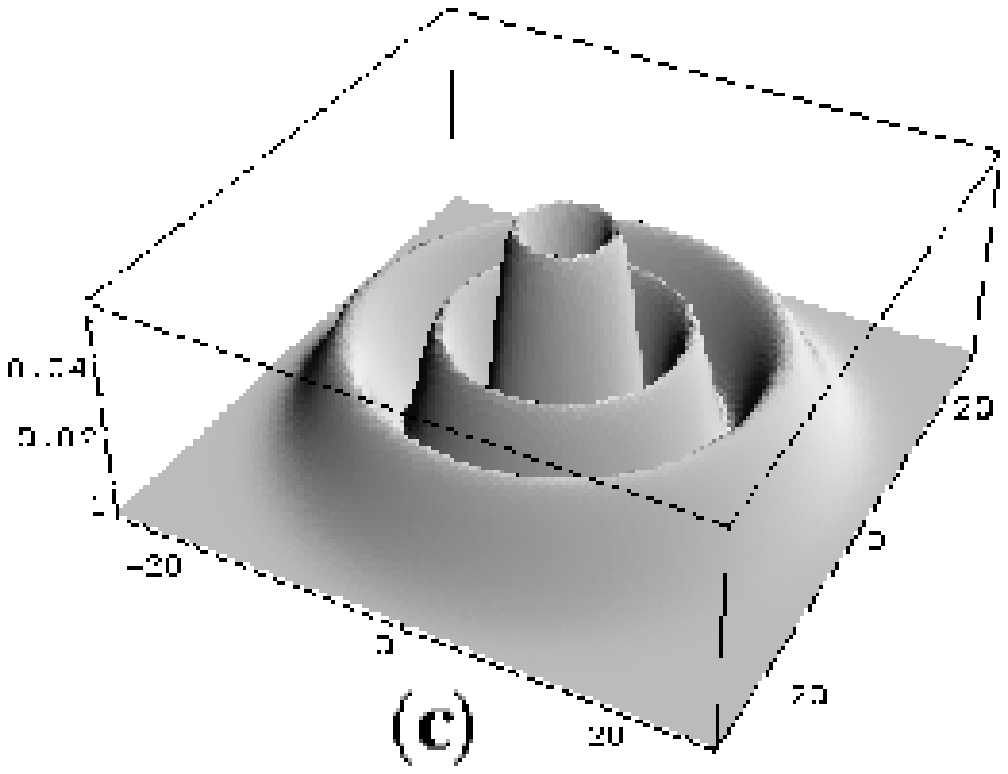}}
\end{tabular}
\caption{\label{SED1}Plots of information entropy density for
attractive $Z=-1$ planar Hooke's atom with $n=3$ and  (a) $m=0$,
(b) $m=1$ and (c) $m=2$.}
\end{figure*}

\begin{figure*}
\begin{tabular}{ccc}
\scalebox{0.45}{\includegraphics{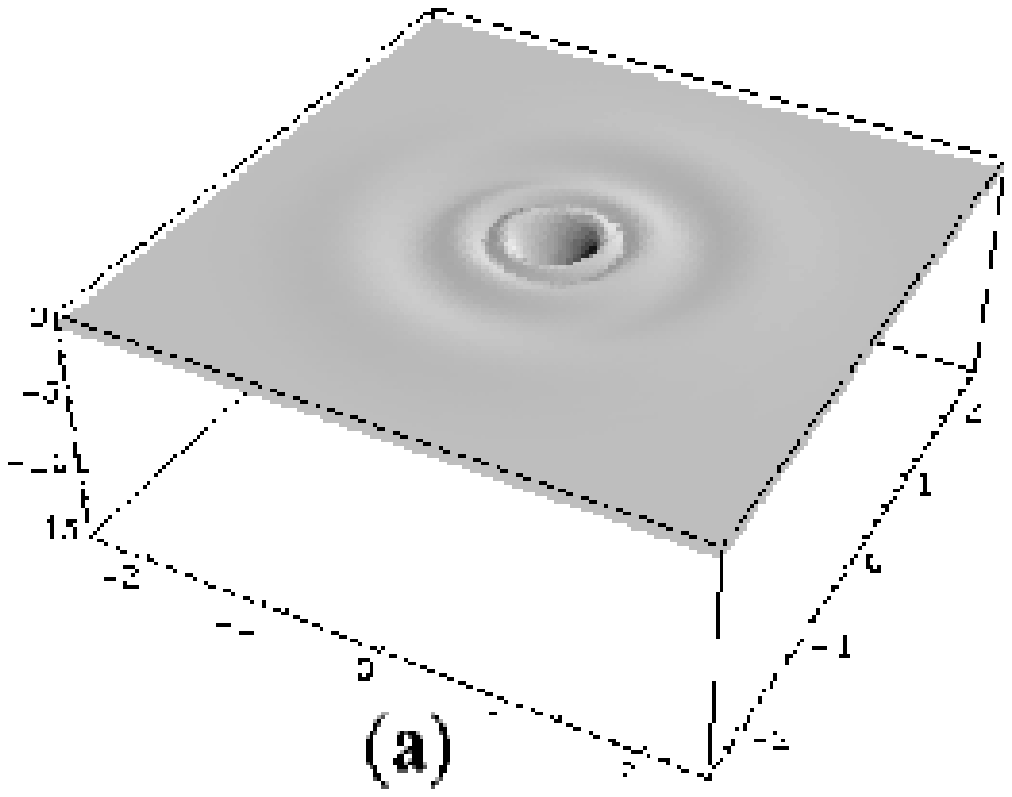}}&
\scalebox{0.45}{\includegraphics{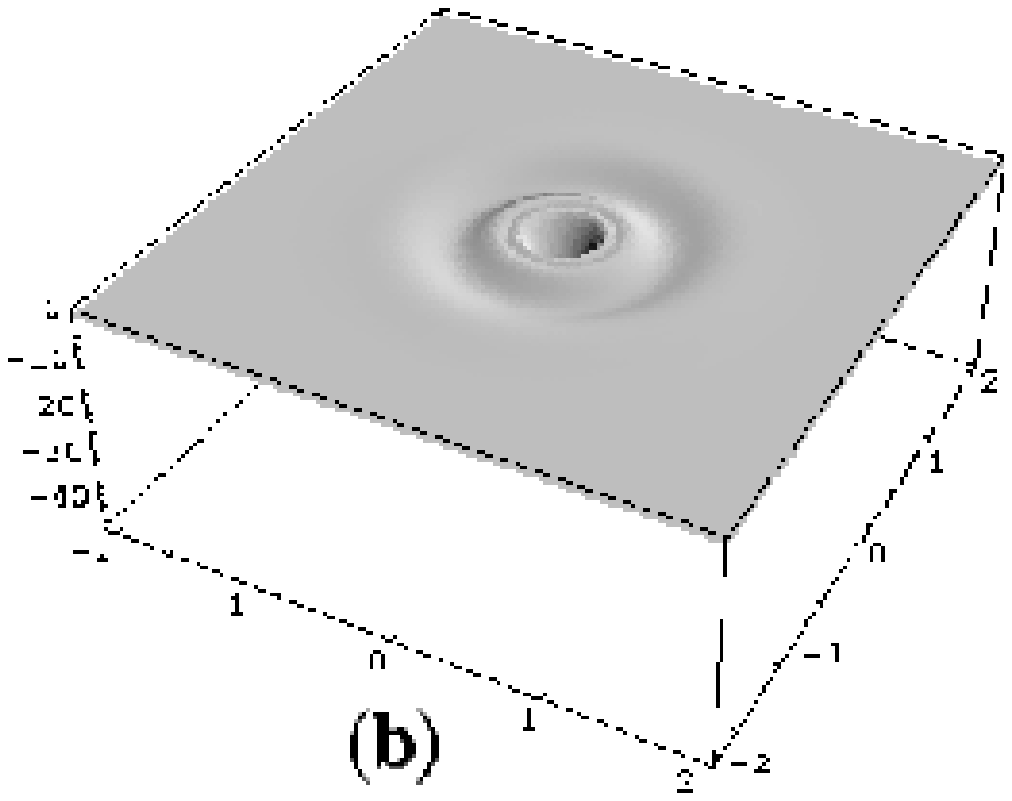}}&
\scalebox{0.45}{\includegraphics{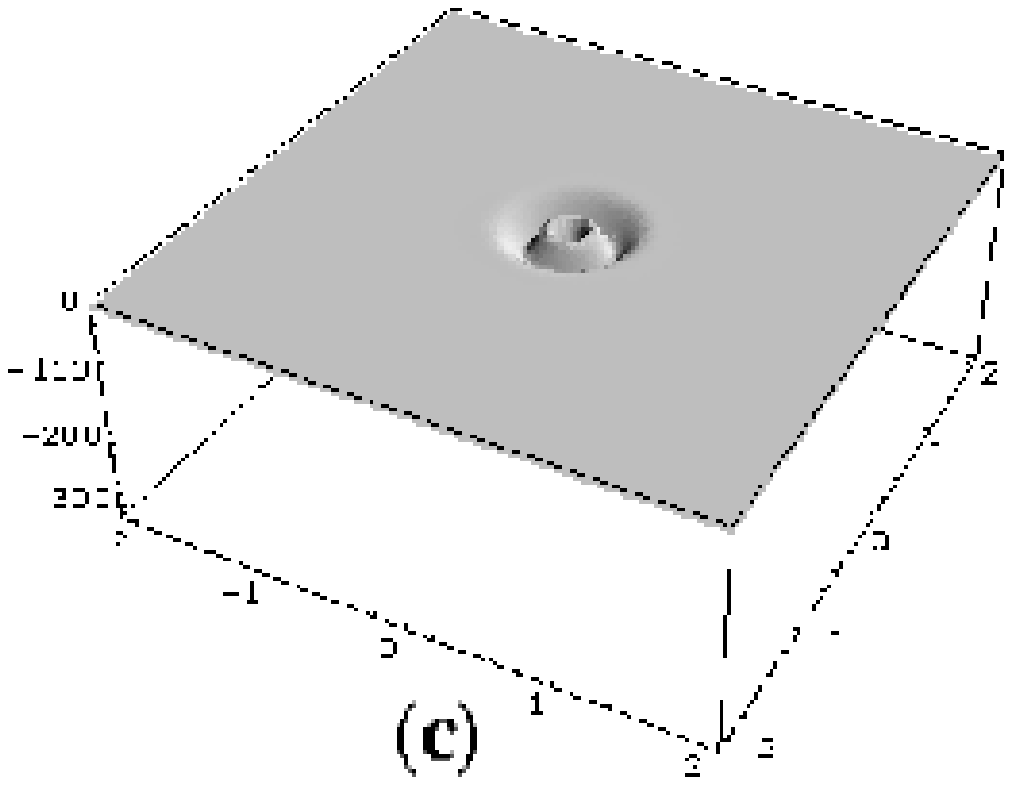}}
\end{tabular}
\caption{\label{SED2}Plots of information entropy density for
attractive planar Hooke's atom with $n=2$ and $m=0$ depicting the
effect of Coulomb coupling strength for (a) $Z=-2$, (b) $Z=-3$ and
(c) $Z=-7$.}
\end{figure*}

\begin{figure*}
\begin{tabular}{ccc}
\scalebox{0.45}{\includegraphics{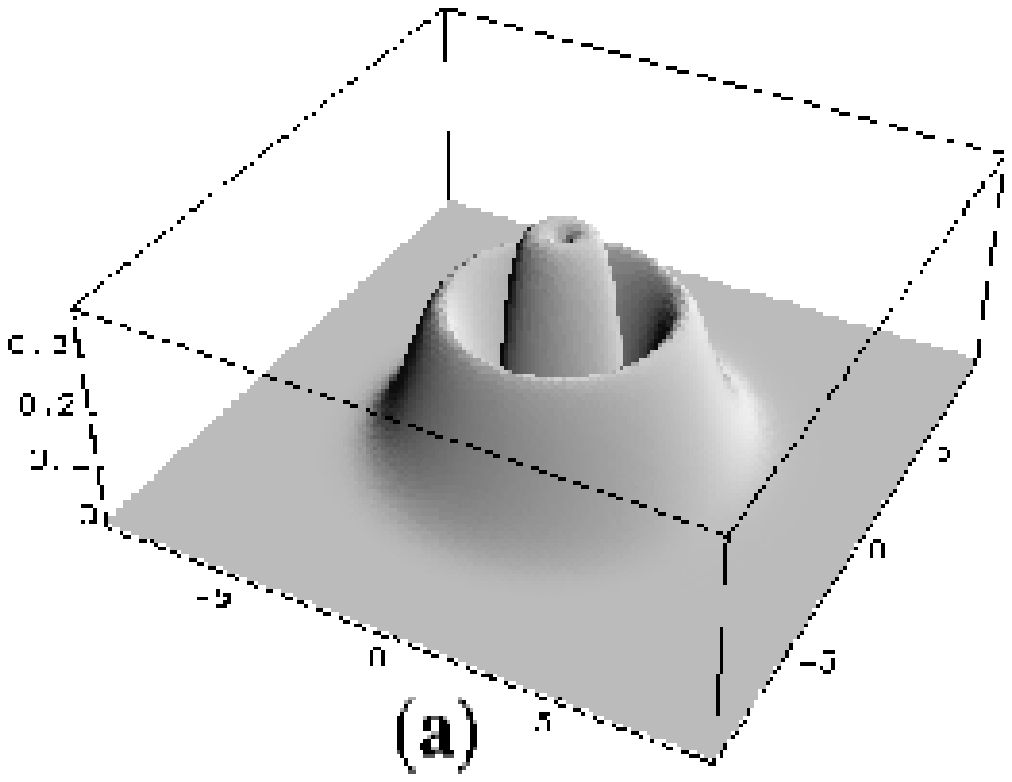}}&
\scalebox{0.45}{\includegraphics{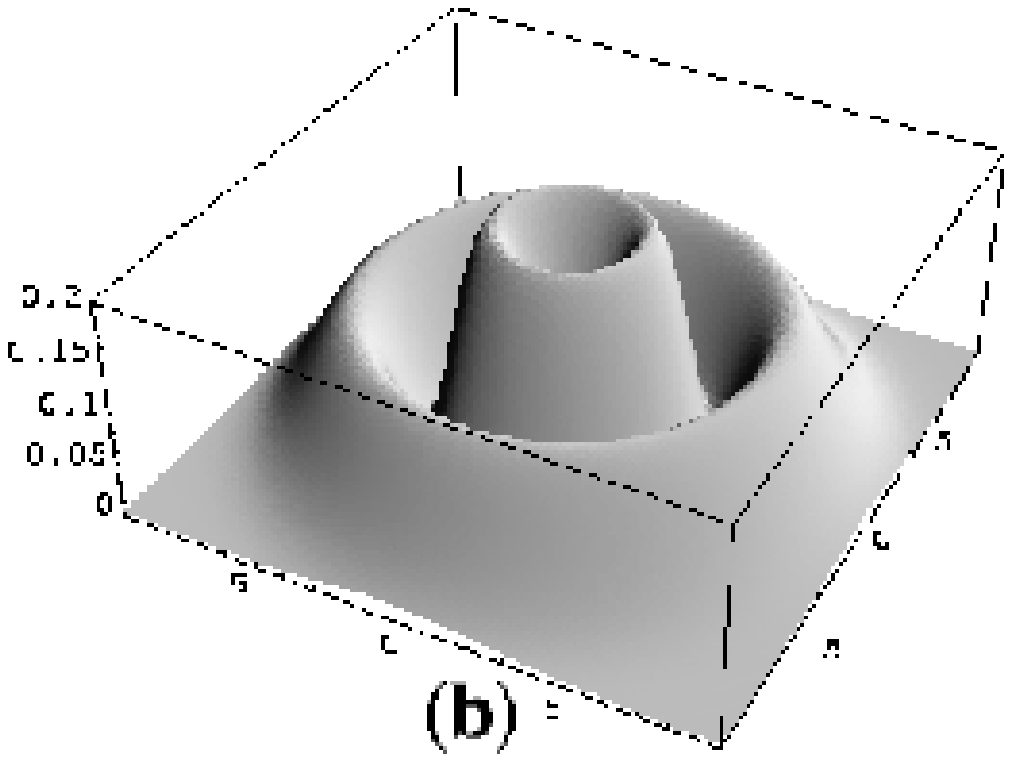}}&
\scalebox{0.45}{\includegraphics{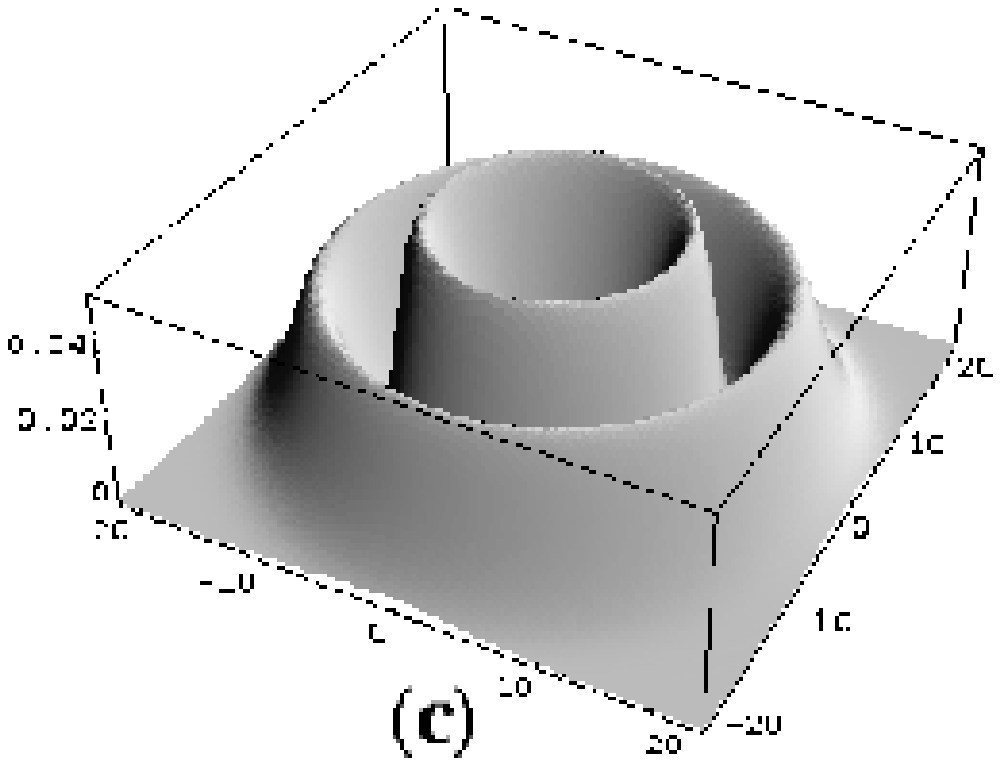}}
\end{tabular}
\caption{\label{SED3}Plots of information entropy density for
repulsive ($Z=1$) planar Hooke's atom with $n=4$ (a) $m=0$, (b)
$m=1$ and (c) $m=5$.}
\end{figure*}

\begin{figure}
\begin{center}
\includegraphics[width=3.0in,clip]{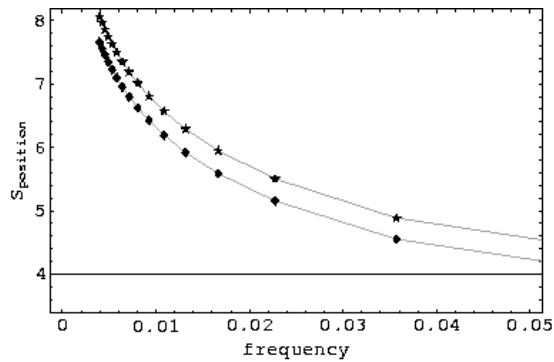}
\caption{\label{labelFig1} Plot depicting the variation of
position space entropy with respect to various allowed oscillator
frequencies for $n=3$, the green line and red lines are for
attractive and repulsive cases respectively.}
\end{center}
\end{figure}

\section{Conclusions} In conclusion, we have illustrated the
utility of a recently developed method for solving linear
differential equations to Hooke's atom The information entropy and
their densities are analyzed systematically for studying the
effect of interaction on correlation. The procedure for developing
perturbative expansions based on the present approach is
indicated. A connection of this dynamical system with well studied
QES systems can also be used for developing a suitable
perturbation theory involving the charge parameter $Z$. The
usefulness of single particle density to density functional theory
is well-known, as also its ability for studying entanglement of
this correlated system. We would like to get back to these
questions in near future. Recently, Ralko and Truong \ct{Truong1}
has connected this system to anyons and Heun's equation \ct{heun}.
This is an interesting direction, needing further investigations.

{\bf{Acknowledgments:}} We acknowledge many useful discussions
with Prof. K.D. Sen, who also brought to our notice many relevant
references. CSM thanks Physical Research Laboratory for the
hospitality during this project.


\end{document}